\documentclass[aip,jmp,amsmath,amssymb,reprint,citeautoscript]{revtex4-2}

\usepackage{graphicx}% Include figure files
\usepackage{dcolumn}% Align table columns on decimal point
\usepackage{bm}% bold math
\newcommand{\aibnaffil}{%
 \affiliation{The Australian Institute for Bioengineering and Nanotechnology, The University of Queensland, St. Lucia, QLD, 4072, Australia}}
 \newcommand{\scmbaffil}{%
 \affiliation{The School of Chemistry and Molecular Biosciences, The University of Queensland, St. Lucia, QLD, 4072, Australia }}
 \newcommand{\getocoaffil}{%
 \affiliation{ARC Centre of Excellence for the Green Electrochemical Transformation of Carbon Dioxide, The University of Queensland, St. Lucia, QLD, 4072, Australia }}

\newcommand{\epsdot}{\dot{\epsilon}}
\newcommand{\gammadot}{\dot{\gamma}}
\newcommand{\gradu}{\nabla\bm{u}}

\begin{document}

\preprint{}

\title{On the importance of numerical integration details for homogeneous flow simulation}

\author{Stephen Sanderson}
 \email{stephen.sanderson@uq.edu.au}
 \aibnaffil
\author{Debra J. Searles}%
 \aibnaffil
 \scmbaffil
 \getocoaffil

\date{\today}% It is always \today, today,

\begin{abstract}
The Sllod equations of motion enable modeling of homogeneous flow at the atomic scale, and are commonly used to predict fluid properties such as viscosity.
However, few publicly available codes support such simulations, and those that do often do not implement a reversible numerical integration scheme or have other subtle problems.
Here, we demonstrate a reversible and energy-conserving integration scheme for the Sllod equations of motion with error on the order of $\delta t^3$, in line with typical operator splitting integrators used in standard molecular dynamics simulations.
We discuss various implementation details, and implement the scheme in LAMMPS where we find that our changes enable more accurate simulation of transient responses, mixed flows, and steady states, especially at high rates of flow.
Importantly, we show that a lack of energy conservation can manifest as a systematic error in the direct ensemble average of the pressure tensor, leading to an error in the calculated viscosity which becomes significant at high flow rates.
\end{abstract}

%Use showkeys class option if keyword
\maketitle

\section{Introduction}

Nonequilibrium molecular dynamics (NEMD) simulations have long proven useful for studying rheological properties of fluids \cite{Evans_1981,Evans_1983,Clarke_1987,Todd_2017,Ewen_2021}.
The Sllod equations of motion \cite{Evans_1984} in particular have proven efficient and effective for the study of bulk properties, as they enable the driving of homogeneous laminar flow in an infinitely periodic unit cell without explicitly modeling walls.
Absent a thermostat (which is required to reach a steady state, but will be discussed later), the Sllod equations of motion are given by
\begin{eqnarray}
    \dot{\bm{q}}_i &=& \frac{\bm{p}_i}{m_i} + \bm{q}_i\cdot\gradu, \label{eqn:qdot} \\
    \dot{\bm{p}}_i &=& \bm{F}_i - \bm{p}_i\cdot\gradu, \label{eqn:pdot} \\
    \dot{\bm{a}} &=& \bm{a}\cdot\gradu,\\
    \dot{\bm{b}} &=& \bm{b}\cdot\gradu,\\
    \dot{\bm{c}} &=& \bm{c}\cdot\gradu,
\end{eqnarray}
where $\bm{q}_i$ is the position of particle $i$, $\bm{p}_i$ is its peculiar momentum (momentum relative to the streaming velocity), $\bm{F}_i$ is the total of the conservative forces acting on the particle, $\bm{a}$, $\bm{b}$ and $\bm{c}$ are the lattice vectors of the unit cell (assumed here to be a general triclinic unit cell), and $\gradu$ is the tensor describing the gradient of the (laminar) flow velocity profile to be imposed, and $\cdot$ represents a tensor contraction. \footnote{The tensor contraction $\bm{u}\cdot\bm{M}$ gives a vector with components $v_j = [\bm{u}\cdot\bm{M}]_j = \sum_i u_i M_{ij}$}
The unperturbed internal energy is
\begin{equation}
\mathcal{H}_0 = \sum_i \frac{\bm{p}_i\cdot\bm{p}_i}{2m_i} + \Phi(\bm{q},\bm{a},\bm{b},\bm{c}),
\end{equation}
where $\Phi(\bm{q}, \bm{a},\bm{b},\bm{c})$ is the potential energy of the particle configuration with the dependence on the lattice vectors of the periodic unit cell made explicit. 
Hence, these equations of motion produce an energy dissipation \cite{EvansMorriss_2007,Todd_1997}
\begin{equation}
\dot{\mathcal{H}}_0 = -V\bm{P}^T:\gradu + \left[\frac{\partial\Phi}{\partial\bm{a}}\bm{a} + \frac{\partial\Phi}{\partial\bm{b}}\bm{b} + \frac{\partial\Phi}{\partial\bm{c}}\bm{c}\right]:\gradu,
\end{equation}
where $V$ is the volume of the unit cell, $\bm{P}$ is the pressure tensor and $:$ is the double contraction. \footnote{The double contraction of two second rank tensors gives a scalar, $\bm{M}:\bm{N}=\sum_{ij}M_{ij}N_{ij}.$}
Importantly, the first term on the right is exactly the energy dissipation expected from hydrodynamics \cite{Daivis_2006}, while the second term has been found to oscillate around zero in the steady state with negligible contribution in the case of large systems with short-ranged interactions \cite{Petravic_1998,Bernardi_2015}.

Viewed from the laboratory frame of reference, and considering the case of the flow being ``turned on'' at time 0 ($\gradu(t) = \gradu\Theta(t)$ with $\Theta(t)$ the Heaviside step function), it follows that
\begin{equation}
m_i\ddot{\bm{q}}_i = \bm{F}_i + m_i\bm{q}_i\cdot\gradu\cdot\gradu\Theta(t) + m_i\bm{q}_i\cdot\gradu\delta(t).
\end{equation}
For shear flow, $\gradu\cdot\gradu = \bm{0}$, and hence the Sllod equations are equivalent in that case to superimposing the expected velocity profile ($\bm{q}_i\cdot\gradu$) and then evolving under Newton's laws, with the added requirement that the periodic unit cell must be evolved in a manner commensurate with the flow.

After initially being implemented and tested in in-house codes, the Sllod equations have for a long time been widely available in LAMMPS, a large-scale, high performance molecular dynamics (MD) package \cite{LAMMPS}.
While other large-scale MD codes do support deformation of the periodic unit cell, they typically implement it in an ad-hoc manner rather than using the Sllod equations of motion, hence LAMMPS is by far the most used implementation for bulk flow modelling.
However, even when the Sllod equations are used, their numerical integration must be performed carefully to avoid subtle errors.

In this manuscript, we discuss various considerations for the numerical integration of the Sllod equations of motion, motivated by small but meaningful errors we found in existing implementations.
The simulation of transient responses,
mixed flows, and steady states, especially at high rates of flow are shown to be more accurate using our new implementation in LAMMPS \cite{LAMMPS}.
We also derive a conserved quantity for the thermostatted Sllod dynamics with arbitrary $\gradu$, which we use as a test for the stability of the integrator.
While numerical integration of the Sllod equations has been considered in the past \cite{Isbister_1997,Zhang_1999,Pan_2005,Mundy_1995}, previous work focused on planar shear flow, whereas here we treat the more general case of arbitrary triangular flow tensors, with some discussion of the completely general case which additionally permits rotational flow.

\section{Conserved quantity \label{sec:conserved-quantity}}

The Sllod equations of motion are in general non-Hamiltonian, but in a similar manner to how a conserved quantity for Nos\'e-Hoover thermostatted dynamics has been obtained,\cite{Martyna_1992} one can write down a conserved quantity for Sllod by introducing an extra phase variable which gives up energy equal to the work done on the fluid.
Such a conserved quantity has previously been shown by Tuckerman et al.\cite{Tuckerman_1997}, in which the kinetic energy term explicitly includes the streaming component of the flow, but here we show a simpler expression is obtained by instead treating the kinetic energy term as the \textit{thermal} kinetic energy.
Additionally, our expression explicitly accounts for the deforming boundary conditions.

We begin by extending phase space with the standard additional degrees of freedom associated with the Nos\'e-Hoover thermostat \cite{Martyna_1992} (a fictitious particle representing the thermal reservoir, with position $\eta$, momentum $p_\eta$ and inertia $Q$).
We then add one extra degree of freedom, $\lambda$, which represents an energy source from which the flow is driven, and the conserved quantity can be written as
\begin{equation}
    \mathcal{H}' = \sum_i \frac{\bm{p}_i\cdot\bm{p}_i}{2m_i} + \Phi(\bm{q},\bm{a},\bm{b},\bm{c}) + \frac{1}{2Q}p_\eta^2 + N_dk_BT\eta + \lambda. \label{eqn:conserved}
\end{equation}
For $\mathcal{H}'$ to be conserved, we require $\dot{\mathcal{H}}'=0$, which dictates the evolution of $\lambda$, giving the equations of motion as
\begin{eqnarray}
    \dot{\bm{q}}_i &=& \frac{\bm{p}_i}{m_i} + \bm{q}_i\cdot\gradu, \nonumber\\
    \dot{\bm{a}} &=& \bm{a}\cdot\gradu, \nonumber\\
    \dot{\bm{b}} &=& \bm{b}\cdot\gradu, \nonumber\\
    \dot{\bm{c}} &=& \bm{c}\cdot\gradu, \nonumber\\
    \dot{\bm{p}}_i &=& \bm{F}_i - \bm{p}_i\cdot\gradu - \frac{p_\eta}{Q}\bm{p}_i, \nonumber\\
    \dot{\eta} &=& \frac{p_\eta}{Q}, \nonumber\\
    \dot{p_\eta} &=& \sum_i \frac{\bm{p}_i\cdot\bm{p}_i}{m_i} - N_dk_BT,\nonumber\\
    \dot{\lambda} &=&  \left[\sum_i \frac{\bm{p}_i\bm{p}_i}{m_i} + \bm{F}_i\bm{q}_i\right]:\gradu -\left[\frac{\partial\Phi}{\partial\bm{a}}\bm{a} + \frac{\partial\Phi}{\partial\bm{b}}\bm{b} + \frac{\partial\Phi}{\partial\bm{c}}\bm{c}\right]:\gradu. \label{eqn:motion-sllod-nh}
\end{eqnarray}
Hence, a stable numerical integration scheme should preserve $\mathcal{H}'$, despite the Sllod equations not being symplectic in general \cite{Searles_1998}.

Note, equivalently, $\eta$ and $\lambda$ can be combined into a single energy reservoir, $\kappa=N_dk_BT\eta + \lambda$, giving
\begin{equation}
    \dot{\kappa} =  N_dk_BT\frac{p_\eta}{Q} + \left[\sum_i \frac{\bm{p}_i\bm{p}_i}{m_i} + \bm{F}_i\bm{q}_i\right]:\gradu -\left[\frac{\partial\Phi}{\partial\bm{a}}\bm{a} + \frac{\partial\Phi}{\partial\bm{b}}\bm{b} + \frac{\partial\Phi}{\partial\bm{c}}\bm{c}\right]:\gradu, 
\end{equation}
with the conserved quantity
\begin{equation}
    \mathcal{H}' = \sum_i \frac{\bm{p}_i\cdot\bm{p}_i}{2m_i} + \Phi(\bm{q},\bm{a},\bm{b},\bm{c}) + \frac{1}{2Q}p_\eta^2 + \kappa. \label{eqn:conserved-single}
\end{equation}
This result can also be formulated in the framework of Sergi et al. \cite{Sergi_2001}, which we show in Appendix \ref{appendix:conserved}.

With $\kappa$ acting as both a source of energy to drive the flow and a sink for energy removed by the thermostat, it reaches a steady state in flows which become steady (i.e. thermostatted, volume-preserving flows).
However, we note that $\langle\dot{\kappa}(t)\rangle$ (and also $\langle\dot{\mathcal{H}}_0(t)\rangle$ and $\langle\dot{p}_\eta(t)\rangle$) may not be instantaneously zero, but periodically oscillate around zero; this is due to the time-periodic nature of the unit cell lattice \cite{Petravic_1998}.
For example, under planar $xy$ shear flow (flow in the x-direction with a gradient in the y direction), the only contributing term is $b_y\gammadot_{xy}\frac{\partial\Phi}{\partial b_x}$, and periodic flipping of the lattice vectors to maintain a bound on the minimum image distance can be viewed as a periodic inversion of $b_y$, hence $\lim_{t\to\infty}\frac{1}{t}\int_{0}^t ds b_y\gammadot_{xy}\frac{\partial\Phi}{\partial b_x} = \text{const.}$ if $\frac{\partial\Phi}{\partial b_x}$ becomes steady or oscillates.
Similarly, evolution of the unit cell under volume-preserving elongational flow can be formulated as a periodic cycle of cell shapes combined with rotation of the lattice with a different period, where the ratio between the two periods is not a rational number \cite{Dobson_2023}, leading to a similar cancellation in the time average at steady state.

Considering $\mathcal{H}_0$ in a steady state, the energy stored in the fluid relative to its energy in the initial (equilibrium) distribution can be evaluated either directly as $\langle\mathcal{H}_0(t)-\mathcal H_0(0)\rangle$, or indirectly by rearranging the conservation condition, $\mathcal{H}'(t) = \mathcal{H}'(0)$, to obtain
\begin{equation}
    \lim_{t\rightarrow\infty}\langle\mathcal{H}_0(t)\rangle-\langle\mathcal{H}_0(0)\rangle = 
        \lim_{t\rightarrow\infty}-\frac{1}{2Q}\bigl(\langle p_\eta(t)\rangle^2 - \langle p_\eta(0)\rangle^2\bigr) - \bigl(\langle\kappa(t)\rangle - \langle\kappa(0)\rangle\bigr),
\end{equation}
where $\langle\cdot\rangle$ denotes an ensemble average.
Note that $\langle p_\eta(0)\rangle=0$ when the initial ensemble is an equilibrium one, and that $\kappa$ does not affect the dynamics and hence $\langle\kappa(0)\rangle$ can be arbitrarily shifted to 0.
Despite $\mathcal{H}_0$, $p_\eta$ and $\kappa$ all reaching a steady state and $\mathcal{H}'$ being conserved, however, the phase space distribution cannot be considered as an equilibrium one, as has been previously demonstrated \cite{Petersen_2022,Evans_2016}.
This is because the equations of motion generate a phase space contraction, $\Lambda = \frac{\partial}{\partial\bm{\Gamma}}\cdot\dot{\bm{\Gamma}} = -3N\frac{p_\eta}{Q} + 3\text{Tr}[\gradu]$, meaning that the phase space probability density function is ever-changing despite low-dimensional observables becoming steady.

\section{Numerical integration \label{sec:coord-integration}}

The Liouville operator for Eqns (\ref{eqn:motion-sllod-nh}) is
\begin{eqnarray}
    i\bm{L} &=& \sum_i \frac{\bm{p}_i}{m_i}\cdot\frac{\partial}{\partial \bm{q}_i}
             + \sum_i \bm{F}_i(\bm{q})\cdot\frac{\partial}{\partial \bm{p}_i} \nonumber\\
            && + \sum_i \bm{q}_i\cdot\gradu\cdot\frac{\partial}{\partial \bm{q}_i}
             - \sum_i \bm{p}_i\cdot\gradu\cdot\frac{\partial}{\partial \bm{p}_i} \nonumber\\
            && + \left[\sum_i \bm{p}_i\bm{p}_i + \bm{F}_i\bm{q}_i j
             - \frac{\partial\Phi}{\partial\bm{a}}\bm{a} - \frac{\partial\Phi}{\partial\bm{b}}\bm{b} - \frac{\partial\Phi}{\partial\bm{c}}\bm{c}
             \right]:\gradu\frac{\partial}{\partial \lambda} \nonumber\\
            && - \frac{p_\eta}{Q}\sum_i\bm{p}_i\cdot\frac{\partial}{\partial\bm{p}_i}
             + \left[\sum_i\frac{\bm{p}_i\cdot\bm{p}_i}{m_i}-N_dk_BT\right]\frac{\partial}{\partial p_\eta}
             + \frac{p_\eta}{Q}\frac{\partial}{\partial \eta} \nonumber\\
             && + \bm{a}\cdot\gradu\cdot\frac{\partial}{\partial\bm{a}}
             + \bm{b}\cdot\gradu\cdot\frac{\partial}{\partial\bm{b}}
             + \bm{c}\cdot\gradu\cdot\frac{\partial}{\partial\bm{c}},
\end{eqnarray}
with the time evolution of a phase point, $\bm{\Gamma}$, described by
\begin{equation}
    \bm{\Gamma}(t) = e^{i\bm{L}t}\bm{\Gamma}(0).
\end{equation}
Applying Trotter factorization to the Liouville operator leads to a reversible, computationally tractable integration scheme in terms of simple, uncoupled differential equations, which is expected to conserve $\mathbf{H}'$ to second order \cite{Martyna_1996}.
To obtain a set of uncoupled differential equations, we first take
\begin{equation}
    i\bm{L} = i\bm{L}_q + i\bm{L}_p + i\bm{L}_{\gradu} + i\bm{L}_\lambda + i\bm{L}_{p_\eta p} + i\bm{L}_{p_\eta} + i\bm{L}_\eta + i\bm{L}_V,
\end{equation}
where
\begin{eqnarray}
    i\bm{L}_q &=& \sum_i \frac{\bm{p}_i}{m_i}\cdot\frac{\partial}{\partial \bm{q}_i}, \nonumber\\
    i\bm{L}_p &=& \sum_i \bm{F}_i(\bm{q})\cdot\frac{\partial}{\partial \bm{p}_i}, \nonumber\\
    i\bm{L}_{\gradu} &=& \sum_i \bm{q}_i\cdot\gradu\cdot\frac{\partial}{\partial \bm{q}_i}
        - \sum_i \bm{p}_i\cdot\gradu\cdot\frac{\partial}{\partial \bm{p}_i}, \nonumber\\
    i\bm{L}_\lambda &=& \left[\sum_i \bm{p}_i\bm{p}_i + \bm{F}_i\bm{q}_i
    -\frac{\partial\Phi}{\partial\bm{a}}\bm{a} - \frac{\partial\Phi}{\partial\bm{b}}\bm{b} - \frac{\partial\Phi}{\partial\bm{c}}\bm{c}\right]:\gradu\frac{\partial}{\partial \lambda}, \nonumber\\
    i\bm{L}_{p_\eta p} &=& - \frac{p_\eta}{Q}\sum_i\bm{p}_i\cdot\frac{\partial}{\partial\bm{p}_i}, \nonumber\\
    i\bm{L}_{p_\eta} &=& \left[\sum_i\frac{\bm{p}_i\cdot\bm{p}_i}{m_i}-N_dk_BT\right]\frac{\partial}{\partial p_\eta}, \nonumber\\
    i\bm{L}_\eta &=& \frac{p_\eta}{Q}\frac{\partial}{\partial \eta}, \nonumber\\
    i\bm{L}_V &=& \bm{a}\cdot\gradu\frac{\partial}{\partial\bm{a}} + \bm{b}\cdot\gradu\frac{\partial}{\partial\bm{b}} + \bm{c}\cdot\gradu\frac{\partial}{\partial\bm{c}}.
\end{eqnarray}
For simple planar shear flow (having a single off-diagonal component to $\gradu$), or for elongational, expanding, or compressing flows (having diagonal $\gradu$), $i\bm{L}_{\gradu}$ and $i\bm{L}_V$ are trivial to apply.
However, under rotational or mixed flows, the Cartesian components become coupled.
For a triangular $\gradu$ (sufficient for all but rotational flows), Appendix \ref{appendix:box-deformation} shows an analytical solution (for the case of $i\bm{L}_V$, but the result may be extended to $i\bm{L}_{\gradu}$).
However, this solution is cumbersome to apply.
Instead, $i\bm{L}_{\gradu}$ (and similarly $i\bm{L}_V$) can be further split into
\begin{eqnarray}
    i\bm{L}_{\gradu_\text{diag}} &=& \sum_i q_{ix} \gradu_{xx}\frac{\partial}{\partial q_{ix}}
        + \sum_i q_{iy} \gradu_{yy}\frac{\partial}{\partial q_{iy}}
        + \sum_i q_{iz} \gradu_{zz}\frac{\partial}{\partial q_{iz}} \nonumber\\
    && -\sum_i p_{ix} \gradu_{xx}\frac{\partial}{\partial p_{ix}}
        - \sum_i p_{iy} \gradu_{yy}\frac{\partial}{\partial p_{iy}}
        - \sum_i p_{iz} \gradu_{zz}\frac{\partial}{\partial p_{iz}}, \nonumber\\
    i\bm{L}_{\gradu_x} &=& \sum_i \left[q_{iy}\gradu_{yx} + q_{iz}\gradu_{zx}\right]\frac{\partial}{\partial q_{ix}} 
        -\sum_i \left[p_{iy}\gradu_{yx} + p_{iz}\gradu_{zx}\right]\frac{\partial}{\partial p_{ix}}, \nonumber\\
    i\bm{L}_{\gradu_y} &=& \sum_i \left[q_{ix}\gradu_{xy} + q_{iz}\gradu_{zy}\right]\frac{\partial}{\partial q_{iy}} 
        -\sum_i \left[p_{ix}\gradu_{xy} + p_{iz}\gradu_{zy}\right]\frac{\partial}{\partial p_{iy}}, \nonumber\\
    i\bm{L}_{\gradu_z} &=& \sum_i \left[q_{ix}\gradu_{xz} + q_{iy}\gradu_{yz}\right]\frac{\partial}{\partial q_{iz}} 
        -\sum_i \left[p_{ix}\gradu_{xz} + p_{iy}\gradu_{yz}\right]\frac{\partial}{\partial p_{iz}}.
\end{eqnarray}
From this, Trotter factorization gives a velocity Verlet-style propagator with the thermostat scheme of Ref. \citenum{Martyna_1996} as
\begin{equation}
\begin{array}{rcll}
    e^{i\bm{L} \delta t} &=&
        % sllod account keeping
        e^{i\bm{L}_\lambda \frac{\delta t}{2}} & \text{(Sllod Bookkeeping)} \\
        % thermostat
        && e^{i\bm{L}_{p_\eta} \frac{\delta t}{4}}
        e^{i\bm{L}_{p_\eta p} \frac{\delta t}{2}}
        e^{i\bm{L}_\eta \frac{\delta t}{2}}
        e^{i\bm{L}_{p_\eta} \frac{\delta t}{4}} & \text{(Thermostat)} \\
        && e^{i\bm{L}_p \frac{\delta t}{2}} & \text{(Velocity)} \\
        && e^{i\bm{L}_{\gradu_\text{diag}} \frac{\delta t}{2}}
        e^{i\bm{L}_{\gradu_z} \frac{\delta t}{2}}
        e^{i\bm{L}_{\gradu_y} \frac{\delta t}{2}}
        e^{i\bm{L}_{\gradu_x} \frac{\delta t}{2}} & \text{(Sllod)} \\
        && e^{i\bm{L}_q \delta t} & \text{(Position)} \\
        && e^{i\bm{L}_V \delta t} & \text{(Unit Cell)} \\
        && e^{i\bm{L}_{\gradu_x} \frac{\delta t}{2}}
        e^{i\bm{L}_{\gradu_y} \frac{\delta t}{2}}
        e^{i\bm{L}_{\gradu_z} \frac{\delta t}{2}}
        e^{i\bm{L}_{\gradu_\text{diag}} \frac{\delta t}{2}} & \text{(Sllod)} \\
        && e^{i\bm{L}_p \frac{\delta t}{2}} & \text{(Velocity)} \\
        && e^{i\bm{L}_{p_\eta} \frac{\delta t}{4}}
        e^{i\bm{L}_\eta \frac{\delta t}{2}}
        e^{i\bm{L}_{p_\eta p} \frac{\delta t}{2}}
        e^{i\bm{L}_{p_\eta} \frac{\delta t}{4}} & \text{(Thermostat)} \\
        && e^{i\bm{L}_\lambda \frac{\delta t}{2}} & \text{(Sllod Bookkeeping)} \\
        && + \mathcal{O}(\delta t^3) &
\end{array},\label{eqn:propagator}
\end{equation}
noting that other choices are also possible, that $i\bm{L}_q$ and $i\bm{L}_V$ commute so they may be applied simultaneously (as is also the case for $i\bm{L}_\eta$ and $i\bm{L}_{p_\eta p}$), and that $\eta$ and $\lambda$ need not be kept track of if their values are not of interest.
Here, the Sllod half-step is performed between the velocity half-step and the position step so that all changes to the position happen consecutively, thereby avoiding situations in which the force would need to be calculated multiple times within a time step.
It is also sometimes convenient to separate $i\bm{L}_{\gradu}$ into position components and momentum components, i.e.
\begin{eqnarray}
    i\bm{L}_{\gradu} &=& i\bm{L}_{q\gradu} + i\bm{L}_{p\gradu},
\end{eqnarray}
\sloppy in which case the same sequence for the position update could be achieved by $e^{i\bm{L}_{q\gradu}\delta t/2} e^{i\bm{L}_q\delta t} e^{i\bm{L}_V\delta t} e^{i\bm{L}_{q\gradu}\delta t/2}$ while applying $i\bm{L}_{p\gradu}$ at a different point in the time step.
In the case of planar shear flow, this scheme reduces to the one previously derived for use with the Nos\'e-Hoover thermostat \cite{Mundy_1995}.

We note that direct integration of the $\left[\frac{\partial\Phi}{\partial\bm{a}}\bm{a} + \frac{\partial\Phi}{\partial\bm{b}}\bm{b} + \frac{\partial\Phi}{\partial\bm{c}}\bm{c}\right]:\gradu$ term in $\dot{\lambda}$ or $\dot{\kappa}$ is numerically difficult due to discontinuities as particles cross periodic boundaries.
However, for short-ranged interactions, the relative contribution of the term shrinks with the system size, and has previously been shown to be negligible even for moderately sized systems \cite{Bernardi_2015}.
As we consider here only short-ranged interactions, we discard that term for simplicity, noting that our tests of varying system sizes showed that finite size effects in the percentage deviation of the conserved quantity in time vanish at 1/4 of the number of particles (and periodic volume) compared to the system sizes used for the figures shown in this work.

\section{Frame of reference\label{sec:reference-frame}}

The Sllod equations of motion are defined in terms of the ``peculiar'' momenta, $\bm{p}_i$, which are the momenta relative to the expected streaming velocity, $\bm{u}(\bm{q}_i) = \bm{q}_i\cdot\gradu$.
For atomic fluids in the laminar flow regime, this corresponds to the thermal momentum, making combination with a thermostat simple (although this can become problematic for more complicated systems or systems in the turbulent regime).
The simplest method for implementing the Sllod equations is to store velocity in the peculiar frame and integrate with a scheme such as Eqn. (\ref{eqn:propagator}), only converting to the lab-frame when needed for the computation of phase variables (e.g. angular velocity).
However, large-scale MD codes often store velocity in the lab-frame for reasons such as the simple modularity and extensibility it offers, and in this case the Sllod equations must be treated with care.
In particular, it is tempting to perform the position update (i.e. $i\bm{L}_q + i\bm{L}_{q\gradu}$) by simply using the lab-frame velocity, since it is equal to $\dot{\bm{q}}_i$.
However, this does not reversibly apply $i\bm{L}_{q\gradu}$, resulting in an effective step change to $\bm{p}_i$ due to the implicit change in streaming velocity from $\bm{q}_i(t)\cdot\gradu$ to $\bm{q}_i(t+\delta t)\cdot\gradu$
This can result in $\mathcal{H}'$ not being conserved, and importantly appears as a systematic error in the pressure (and therefore in $\dot{\lambda}$), as we show in Section \ref{sec:results}.
The problem can be worked around by, during the position update, first converting velocity to the peculiar frame, updating the position using Eqn. (\ref{eqn:qdot}), and then converting velocity back to the lab-frame using the new position.
Additionally, care must be taken to adjust lab-frame velocities whenever particles are remapped to a new unit cell, and lattice vectors must be up-to-date when used with fractional particle coordinates to calculate streaming velocity.

\section{Boundary integration}

For planar shear flow, Lees-Edwards (sliding brick) boundary conditions provide a simple method for representing the deformation of the unit cell \cite{Lees_1972}, but elongational flows and other more complicated flows are easiest to treat using a triclinic unit cell, which facilitates advanced algorithms needed for simulations of arbitrary duration \cite{Hunt_2003,Hunt_2010,Dobson_2014,Hunt_2015,Dobson_2023}.
Eqn. (\ref{eqn:motion-sllod-nh}) shows that each lattice vector of the unit cell can simply be integrated as if it were a non-interacting point particle with zero peculiar momentum, which can be done either analytically or by splitting $i\bm{L}_V$ as described in Section \ref{sec:coord-integration}.
An analytical solution in theory allows for the box shape on a particular time step to be solved for directly from its initial shape at $t=0$, as is implemented in LAMMPS, for example, but care must be taken to correctly handle couplings between components of $\gradu$, which LAMMPS does not currently account for.
Furthermore, the direct analytical solution is made more difficult by the fact that the unit cell is remapped onto an equivalent lattice to preserve a lower-bound on the minimum interaction distance, so this remapping must be performed each time after solving for the analytical box shape.
Additionally, under mixed shear flow with both an $xy$ and a $yz$ component, a continuous change is induced in the $xz$ tilt after the unit cell has been remapped, which must be handled carefully.
Hence, integrating the lattice vectors directly as part of each time step may be preferable to avoid these difficulties.

In either case, it is clear from Eqn. (\ref{eqn:propagator}) that the update to the unit cell should occur in tandem with the update to particle positions to avoid multiple force calculations.
If, for example, the box is updated at the end of each full step, but force is only calculated after the position update, then each force calculation will use a box shape which is $\frac{1}{2}\delta t$ behind where it should be.
While this has minimal impact in the steady state, it can be important for the transient response.
The unit cell should also be updated every time step (rather than once every $N > 1$ steps) to avoid inducing unphysical stresses across the periodic boundaries.

\section{Results \label{sec:results}}

To test the integration scheme described in this work, we have made a number of modifications to LAMMPS, in particular to \texttt{fix nvt/sllod} and \texttt{fix deform}, in order for $\mathcal{H}'$ to be conserved.
They were to:
\begin{itemize}
    \item Adjust the integration steps to match Eqn. (\ref{eqn:propagator}).
    \item Allow unit cell deformation to occur each time step, immediately after the position update and before force calculation.
    \item Account for mixed flows in the unit cell update, as per Appendix \ref{appendix:box-deformation}. This included correcting the calculation of the rate of change of the unit cell, which was required for correct remapping of lab-frame velocities and determination of $\gradu$ from the box shape.
    \item Reversibly integrate lab-frame velocities as described in Section \ref{sec:reference-frame}.
    \item Fix a bug where particle lab-frame velocities were not correctly adjusted when the unit cell was remapped onto an equivalent lattice.
\end{itemize}
We also added support for storing velocity in the peculiar frame, which is more computationally efficient, and our changes allow long-running simulations with changes in both the $xy$ and $yz$ tilt to occur by correctly accounting for the $xz$ component after the unit cell is remapped.
In the following figures, we compare the current LAMMPS implementation (version \texttt{stable\_29Aug2024\_update4}), labeled `LMP', and our modified version with velocity stored in the lab- and peculiar-frames, labeled `Lab-Frame' and `Peculiar', respectively.
Note, we focus on \texttt{fix nvt/sllod} here, and not the alternative implementation in the UEF package \cite{Nicholson_2016}, which treats velocity in the peculiar frame and correctly integrates the unit cell with the particle positions, but is compatible only with traceless diagonal $\gradu$ and therefore does not need to handle the complexities of couplings between directions.

Unless otherwise stated, data was generated as follows, and all quantities are in Lennard-Jones reduced units.
50 independent systems were generated with different random initial velocities, and equilibrated for 25 normalized time units.
Configurations were then sampled from each system at intervals of 2.5 time units to collect a total of 10,000 independent configurations from the equilibrium ensemble.
To each configuration, four mappings were applied to generate distinct, but equally probable configurations.
These were
\begin{eqnarray}
    Id. &=& \{q_x, q_y, q_z, p_x, p_y, p_z\} \rightarrow \{ q_x, q_y, q_z, p_x, p_y, p_z\}, \\
    \mathcal{M}_{p} &=& \{q_x, q_y, q_z, p_x, p_y, p_z\} \rightarrow \{ q_x, q_y, q_z, -p_x, -p_y, -p_z\}, \\
    \mathcal{M}_{x} &=& \{q_x, q_y, q_z, p_x, p_y, p_z\} \rightarrow \{-q_x, q_y, q_z, -p_x, p_y, p_z\}, \\
    \mathcal{M}_{x,p} &=& \{q_x, q_y, q_z, p_x, p_y, p_z\} \rightarrow \{-q_x, q_y, q_z, p_x, -p_y, -p_z\}.
\end{eqnarray}
This produced a total of 40,000 initial conditions to which the Sllod equations of motion were applied with various $\gradu$ of increasing complexity.
A consistent integration time step ($\delta t = 0.001$ unless specified otherwise) was used throughout equilibration and nonequilibrium simulations, and neighbor lists were calculated each time step to ensure reversibility.
The Nos\'e-Hoover thermostat had a coupling time of $\tau=0.25$ ($Q = N_dk_BT\tau^2$), and the values of the extended phase space variables associated with the thermostat were carried through from the equilibrium snapshots into the nonequilibrium simulations.
Error bars and error envelopes represent one standard error in the mean.
To correctly determine the statistical uncertainty in the result, the trajectories generated by each map applied to a given phase point were first averaged together into a single data point, and then the standard error was calculated from those 10,000 data points.

\begin{figure}
    \centering
    \includegraphics[width=\linewidth]{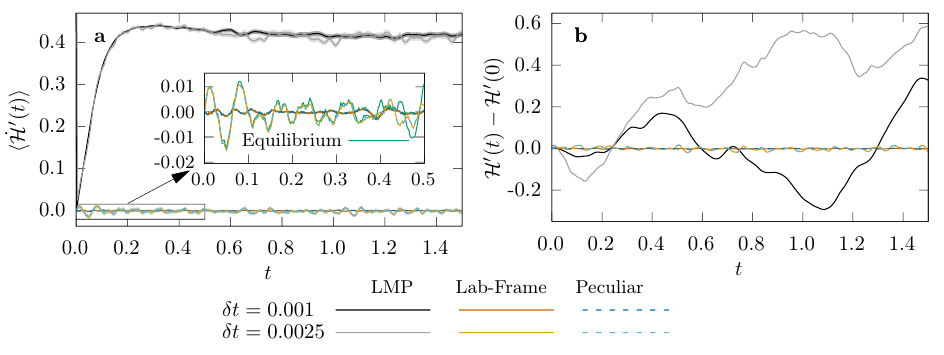}
    \caption{
    Check for conservation of $\mathcal{H}'$ under planar shear flow ($\gammadot_{xy} = 0.1$) with two different integration time steps, showing (a) the rate of change in the conserved quantity averaged over 40,000 trajectories and (b) the conserved quantity for a single trajectory.
    Note, the data from the peculiar and lab-frame implementations overlap.
    The inset in (a) additionally shows the rate of change in the conserved quantity at equilibrium for comparison (error envelopes omitted for clarity).
    }
    \label{fig:xy-conservation}
\end{figure}

\begin{figure}
    \centering
    \includegraphics[width=0.5\linewidth]{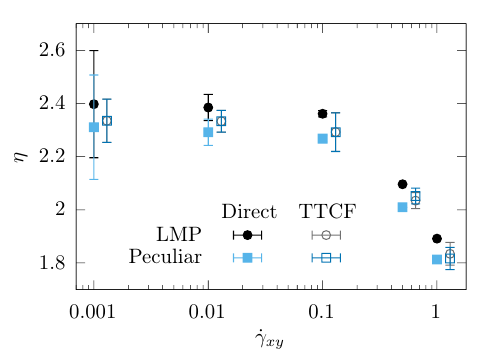}
    \caption{
    Comparison of the shear rate dependence of the shear viscosity, $\eta = {-\langle P_{xy}\rangle}/{\gammadot_{xy}}$, using the time- and ensemble-averaged shear pressure over the final 0.5 time units of each simulation as given by the direct average (closed symbols) and the TTCF formalism (open symbols).
    TTCF results are offset on the $x$ axis for clarity, but correspond to the same shear rate values as those of the direct average results.
    To reduce uncertainty, 200,000 trajectories were used for $\gammadot_{xy}\in\{0.001, 0.5, 1\}$ rather than 40,000.
    To ensure sampling of the steady state, a trajectory length of 2.5 time units was used for $\gammadot_{xy} = 1$ rather than 1.5 as used for other simulations.
    Results from the lab-frame implementation overlapped exactly with those of the peculiar frame implementation, and were therefore omitted for clarity.
    }
    \label{fig:viscosity}
\end{figure}

Figure \ref{fig:xy-conservation} shows the response of the conserved quantity, $\mathcal{H}'$, to planar $xy$ shear flow with $\gammadot_{xy} = 0.1$, calculated as a direct ensemble average, for a system of 256 Weeks-Chandler-Anderson \cite{Weeks_1971} particles in an initially cubic unit cell with density 0.8442 and a normalized temperature of 0.722.
It is clear that both the peculiar-frame and lab-frame implementations conserve $\mathcal{H}'$ to the same degree as it is conserved in an equilibrium simulation, both on average and for an individual simulation, whereas the `LMP' implementation does not.
Beyond the systematic error in the steady state, caused by the irreversibility of the integration scheme, the value of $\dot{\mathcal{H}}'(0)$ is also very large due to the unit cell not being updated in-time with the particle positions, but instead being updated at the end of the step.
Importantly, Fig. \ref{fig:viscosity} shows that the lack of conservation of $\dot{\mathcal{H}}'$ manifests as a larger magnitude of the shear pressure, which leads to an overestimation of the shear viscosity when calculated using the steady state value of $\langle P_{xy} \rangle$, especially at higher shear rates.
Interestingly, application of exact response theory in the form of the Transient Time Correlation Function (TTCF) formalism to calculate viscosity gives good agreement between all three integration schemes, in alignment with the direct average results of the energy-conserving schemes.
We expect this is due to the fact that TTCF calculates properties from the fluctuations, rather than a direct average, and so the systematic error present in the direct average does not appear since it only enters after the system begins responding (i.e. the initial value of $P_{xy}$ is correct) and it does not significantly affect the fluctuations, only the value about which they fluctuate.
Note, this explains the discrepancy between direct average and TTCF results which has previously been observed at shear rates similar to those shown in Fig. \ref{fig:viscosity} \cite{Maffioli_2024}.

\begin{figure}
    \centering
    \includegraphics[width=\linewidth]{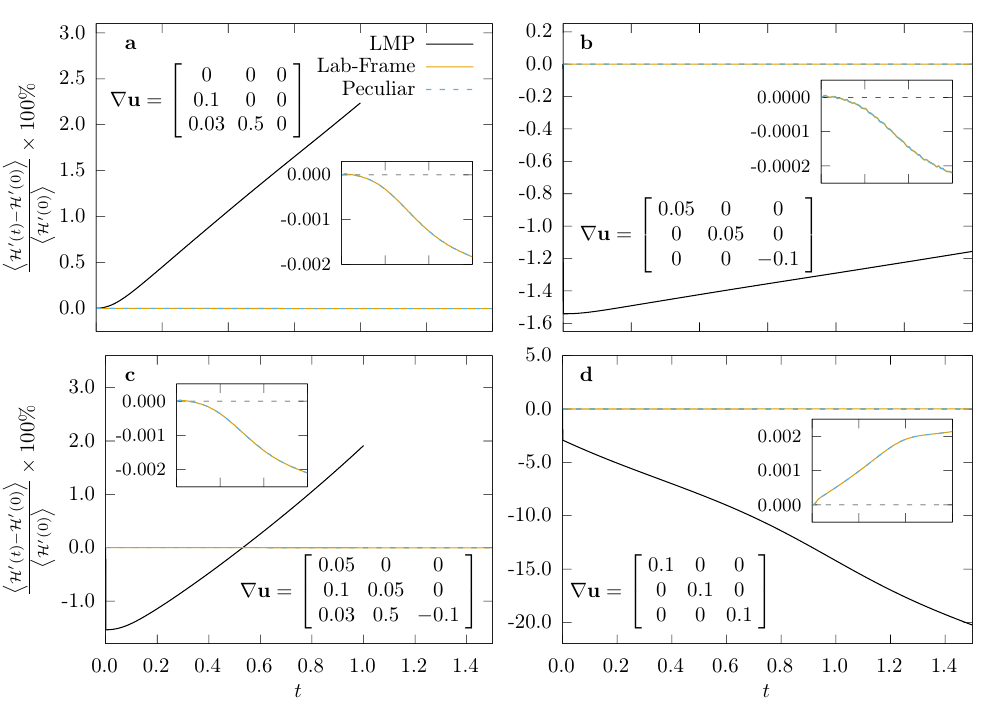}
    \caption{
     Average percentage change in the conserved quantity under (a) mixed shear flow ($\gammadot_{xy} = 0.1$, $\gammadot_{xz} = 0.03$ and $\gammadot_{yz} = 0.5$), (b) biaxial extensional flow ($\epsdot_{xx} = \epsdot_{yy} = -\frac{1}{2}\epsdot_{zz} = 0.05$), (c) mixed biaxial extensional and shear flow ($\epsdot_{xx}=\epsdot_{yy}=-\frac{1}{2}\epsdot_{zz} = 0.05$, $\gammadot_{xy} = 0.1$, $\gammadot_{xz} = 0.03$, $\gammadot_{yz} = 0.5$), and (d) uniform expansion ($\epsdot_{xx}=\epsdot_{yy}=\epsdot_{zz}=0.1$).
    The insets show only the `Lab-Frame' and `Peculiar' curves, which all remain within approximately $0.002\%$ change in $\mathcal{H}'$ over the duration simulated.
    Note, the `LMP' implementation does not allow for remapping of the $yz$ tilt under the flows of (a) or (c), and hence the simulation time was reduced in those cases.
    The non-zero initial values of the `LMP` curves with extensional flows are caused by a bug which results in the particles being integrated with $\epsdot_{xx} = \epsdot_{yy} = \epsdot_{zz} = 0$ in the first time step.
    }
    \label{fig:mixed-flows}
\end{figure}

To test the handling of interaction between shear in different directions, we next examine the energy conservation under mixed shear flow with $\gammadot_{xy} = 0.1$, $\gammadot_{xz} = 0.03$, and $\gammadot_{yz} = 0.5$.
Figure \ref{fig:mixed-flows}a shows that $\mathcal{H}'$ is conserved under both our peculiar and lab-frame schemes, even through multiple remappings of the unit cell, whereas the `LMP' implementation did not support remapping of the $yz$ tilt, and did not conserve $\mathcal{H}'$ even before that.
Similar results can be seen in Figures \ref{fig:mixed-flows}b-d for biaxial extensional flow, uniformly expanding flow, and a combination of mixed shear with biaxial extension (the most complicated flow supported by a reduced triclinic unit cell, having a fully populated triangular $\gradu$).

\begin{figure}
    \centering
    \includegraphics[width=\linewidth]{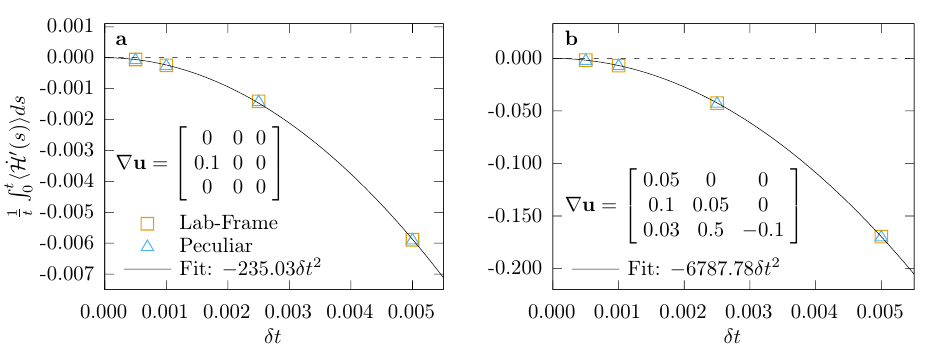}
    \caption{
    Ensemble average rate of change in the conserved quantity, $\langle\dot{\mathcal{H}}'\rangle$, averaged over the duration of the simulation, plotted as a function of the integration time step for (a) planar shear flow and (b) mixed shear and biaxial extensional flow.
    The solid line shows a least squares fit of the form $a\delta t^2$.
    }
    \label{fig:deviation-vs-timestep}
\end{figure}

Lastly, for our energy-conserving implementations, Fig. \ref{fig:deviation-vs-timestep} plots $\dot{\mathcal{H}}'$ averaged in time and across trajectories as a function of the integration time step, where the kinetic and potential energy contributions to $\dot{\mathcal{H}}'$ were calculated using a 2\textsuperscript{nd} order finite difference scheme, while the remaining terms were calculated directly.
It is clear that $\dot{\mathcal{H}}' = 0$ in the limit that $\delta t\to 0$, and that it scales as $\mathcal{O}(\delta t^2)$, as expected from Eqn. (\ref{eqn:propagator}) which has an $\mathcal{O}(\delta t^3)$ error in the propagated quantity.
This is true regardless of the complexity of the simulated flow.

\section{Conclusion}

In summary, we have demonstrated a reversible numerical integration scheme for Sllod dynamics which we have shown to be energy-conserving for general triangular flow tensors.
The scheme was implemented and tested in LAMMPS, where it was found to enable more accurate simulation of transient responses and mixed flows, and more accurate calculation of viscosity from steady states under high shear rates.
Importantly, we find that our improvements to the integration scheme lead to a decrease in the direct average of the viscosity at high flow rates, bringing it in line with calculations based in exact response theory where a discrepancy has previously been reported.

\begin{acknowledgments}
The authors thank the Australian Research Council for its support for this project through the Discovery program (FL190100080).
We acknowledge access to computational resources provided by the Pawsey Supercomputing Centre with funding from the Australian Government and the government of Western Australia and support through the Pawsey Centre for Extreme Scale Readiness (PaCER) scheme.
We also thank Billy Todd and Peter Daivis for their helpful insight and support, and Pieter in 't Veld for his kind comments on a preliminary version of this manuscript.

\end{acknowledgments}

\section*{Author Declarations}
\subsection*{Conflict of Interest}
The authors have no conflicts to disclose.

\subsection*{Author Contributions}
\textbf{Stephen Sanderson:} Conceptualization; Data curation; Formal analysis; Investigation; Methodology; Software; Validation; Visualization; Writing - original draft preparation; Writing - review \& editing (equal).
\textbf{Debra J. Searles:} Funding acquisition; Project administration; Resources; Supervision; Writing - review \& editing (equal).

\section*{Data Availability}
The data that support the findings of this study are available from the corresponding author upon reasonable request.

\appendix

\section{Alternative derivation of conserved quantity\label{appendix:conserved}}

As an alternate route to obtaining equations of motion with a conserved quantity, one can consider that for $\mathcal{H}'$ to be conserved, the dynamics should be given by\cite{Sergi_2001}
\begin{equation}
    \dot{\bm{\Gamma}} = \bm{B}\frac{\partial\mathcal{H}'}{\partial\bm{\Gamma}},
\end{equation}
where $\bm{B}$ is an antisymmetric matrix ($\bm{B}^T=-\bm{B}$) which may depend on the particular point in phase space.
This is because $\dot{\mathcal{H}}'=\frac{1}{2}(\dot{\bm{\Gamma}}\cdot\frac{\partial\mathcal{H}'}{\partial\bm{\Gamma}}+\frac{\partial\mathcal{H}'}{\partial\bm{\Gamma}}\cdot\dot{\bm{\Gamma}})=\frac{1}{2}(\frac{\partial\mathcal{H}'}{\partial\bm{\Gamma}}^T\bm{B}^T\frac{\partial\mathcal{H}'}{\partial\bm{\Gamma}}+\frac{\partial\mathcal{H}'}{\partial\bm{\Gamma}}^T\bm{B}\frac{\partial\mathcal{H}'}{\partial\bm{\Gamma}})=0$ when $\bm{B}^T=-\bm{B}$.
To generate the Nos\'e-Hoover thermostatted Sllod equations of motion with a single energy reservoir, $\kappa$, and conserved $\mathcal{H}'$ as given in Eqn. (\ref{eqn:conserved-single}), we find
\begin{equation}
    \renewcommand{\arraystretch}{1.4}
    \left[\begin{array}{c}
    \dot{\bm{q}}_i \\
    \dot{\bm{a}} \\
    \dot{\bm{b}} \\
    \dot{\bm{c}} \\
    \dot{\kappa} \\
    \dot{\bm{p}}_i \\
    \dot{p_\eta}
    \end{array}\right]
    =
    \left[\begin{array}{ccccccc}
    \bm{0} & \bm{0} & \bm{0} & \bm{0} & \bm{q}_i\cdot\gradu & \bm{1} & \bm{0} \\
    \bm{0} & \bm{0} & \bm{0} & \bm{0} & \bm{a}\cdot\gradu & \bm{0} & \bm{0} \\
    \bm{0} & \bm{0} & \bm{0} & \bm{0} & \bm{b}\cdot\gradu & \bm{0} & \bm{0} \\
    \bm{0} & \bm{0} & \bm{0} & \bm{0} & \bm{c}\cdot\gradu & \bm{0} & \bm{0} \\
    -\bm{q}_i\cdot\gradu & -\bm{a}\cdot\gradu & -\bm{b}\cdot\gradu & -\bm{c}\cdot\gradu & 0 & \bm{p}_i\cdot\gradu & N_dk_BT \\
    -\bm{1} & \bm{0} & \bm{0} & \bm{0} & -\bm{p}_i\cdot\gradu & \bm{0} & -\bm{p}_i \\
    \bm{0} & \bm{0} & \bm{0} & \bm{0} & -N_dk_BT & \bm{p}_i & 0
    \end{array}\right]
    \left[\begin{array}{c}
    \frac{\partial\Phi}{\partial\bm{q}_i} \\
    \frac{\partial\Phi}{\partial\bm{a}} \\
    \frac{\partial\Phi}{\partial\bm{b}} \\
    \frac{\partial\Phi}{\partial\bm{c}} \\
    1 \\
    \frac{\bm{p}_i}{m_i} \\
    \frac{p_\eta}{Q}
    \end{array}\right],
\end{equation}
Here, it is clear that inclusion of $\kappa$ in $\mathcal{H}'$ enables great freedom in the construction of the equations of motion while maintaining a conserved quantity.
Since $\frac{\partial\mathcal{H}'}{\partial\kappa} = 1$, arbitrary terms can be added to the evolution of other phase variables without the value of $\kappa$ influencing the rest of the system, while $\kappa$ keeps account of any energy added or removed by those same terms.

\section{Analytical solution to unit cell deformation}\label{appendix:box-deformation}

For a triangular flow tensor given by
\begin{equation}
\gradu = \left[\begin{array}{c c c}
\epsdot_{xx} & 0 & 0 \\
\gammadot_{xy} & \epsdot_{yy} & 0 \\
\gammadot_{xz} & \gammadot_{yz} & \epsdot_{zz}
\end{array}\right],
\end{equation}
the shape of a reduced triclinic simulation box (in the absence of any remapping) can be solved analytically by integrating the three basis vectors
\begin{eqnarray}
\mathbf{a} &= \left[\begin{array}{c}
h_{xx} \\ 0 \\ 0
\end{array}\right], \\
\mathbf{b} &= \left[\begin{array}{c}
h_{xy} \\ h_{yy} \\ 0
\end{array}\right], \\
\mathbf{c} &= \left[\begin{array}{c}
h_{xz} \\ h_{yz} \\ h_{zz}
\end{array}\right].
\end{eqnarray}
Eqn. (\ref{eqn:motion-sllod-nh}) gives
\begin{eqnarray}
\dot{\mathbf{a}} &=& \mathbf{a}\cdot\gradu, \\
\dot{\mathbf{b}} &=& \mathbf{b}\cdot\gradu, \\
\dot{\mathbf{c}} &=& \mathbf{c}\cdot\gradu,
\end{eqnarray}
resulting in six equations in total for the six non-zero elements of the basis vectors
\begin{eqnarray}
\dot{h}_{xx} &=& \epsdot_{xx}h_{xx}, \label{h_xx_ode}\\
\dot{h}_{yy} &=& \epsdot_{yy}h_{yy}, \label{h_yy_ode}\\
\dot{h}_{zz} &=& \epsdot_{zz}h_{zz}, \label{h_zz_ode}\\
\dot{h}_{xy} &=& \epsdot_{xx}h_{xy} + \gammadot_{xy}h_{yy}, \label{h_xy_ode}\\
\dot{h}_{xz} &=& \epsdot_{xx}h_{xz} + \gammadot_{xz}h_{zz} + \gammadot_{xy}h_{yz}, \label{h_xz_ode}\\
\dot{h}_{yz} &=& \epsdot_{yy}h_{yz} + \gammadot_{yz}h_{zz}. \label{h_yz_ode}
\end{eqnarray}
Solving these equations results in deformation of the periodic unit cell in a manner which is commensurate with the applied flow profile, $\gradu$.
Note that integration of the $\bm{c}$ vector could instead be viewed as application of $i\bm{L}_{q\gradu}$ to a particle's position (or application of $i\bm{L}_{p\gradu}$ to a particle's momentum), and hence this solution could also be applied to particle integration in order to avoid further splitting of $i\bm{L}_{\gradu}$.
Eqns (\ref{h_xx_ode}-\ref{h_zz_ode}) have the simple solutions of
\begin{eqnarray}
h_{xx}(t) &=& h_{xx}(0)e^{\epsdot_{xx}t}, \\
h_{yy}(t) &=& h_{yy}(0)e^{\epsdot_{yy}t}, \\
h_{zz}(t) &=& h_{zz}(0)e^{\epsdot_{zz}t}.
\end{eqnarray}
Eqns (\ref{h_xy_ode}) and (\ref{h_yz_ode}) take the same form as each other, and both have five different cases depending on the values of the diagonal elements of the flow tensor:
\begin{widetext}
\begin{equation}
h_{xy}(t) = h_{xy}(0)e^{\epsdot_{xx}t} + \left\{\begin{array}{l l}
\gammadot_{xy}h_{yy}(0)t, & \epsdot_{xx} = \epsdot_{yy} = 0,\\
\vspace{1.5ex}\\
h_{yy}(0)\frac{\gammadot_{xy}}{\epsdot_{yy}}\left(e^{\epsdot_{yy}t}-1\right), & \epsdot_{xx} = 0, \epsdot_{yy}\neq 0, \\
\vspace{1.5ex}\\
h_{yy}(0)\frac{\gammadot_{xy}}{\epsdot_{xx}}\left(e^{\epsdot_{xx}t}-1\right), & \epsdot_{xx}\neq 0, \epsdot_{yy}=0, \\
\vspace{1.5ex}\\
h_{yy}(0)\gammadot_{xy}t e^{\epsdot_{xx}t}, & \epsdot_{xx}\neq 0, \epsdot_{yy} \neq 0, \epsdot_{xx}=\epsdot_{yy} \\
\vspace{1.5ex}\\
h_{yy}(0)\frac{\gammadot_{xy}}{\epsdot_{yy}-\epsdot_{xx}}\left(e^{\epsdot_{yy}t}-e^{\epsdot_{xx}t}\right), & \epsdot_{xx}\neq 0, \epsdot_{yy}\neq 0, \epsdot_{xx}\neq\epsdot_{yy}
\end{array}\right.
\end{equation}
\begin{equation}
h_{yz}(t) = h_{yz}(0)e^{\epsdot_{yy}t} + \left\{\begin{array}{l l}
\gammadot_{yz}h_{zz}(0)t, & \epsdot_{yy} = \epsdot_{zz} = 0,\\
\vspace{1.5ex}\\
h_{zz}(0)\frac{\gammadot_{yz}}{\epsdot_{zz}}\left(e^{\epsdot_{zz}t}-1\right), & \epsdot_{yy} = 0, \epsdot_{zz}\neq 0, \\
\vspace{1.5ex}\\
h_{zz}(0)\frac{\gammadot_{yz}}{\epsdot_{yy}}\left(e^{\epsdot_{yy}t}-1\right), & \epsdot_{yy}\neq 0, \epsdot_{zz} = 0, \\
\vspace{1.5ex}\\
h_{zz}(0)\gammadot_{yz}t e^{\epsdot_{yy}t}, & \epsdot_{yy}\neq 0, \epsdot_{zz} \neq 0, \epsdot_{yy}=\epsdot_{zz} \\
\vspace{1.5ex}\\
h_{zz}(0)\frac{\gammadot_{yz}}{\epsdot_{zz}-\epsdot_{yy}}\left(e^{\epsdot_{zz}t}-e^{\epsdot_{yy}t}\right), & \epsdot_{yy}\neq 0, \epsdot_{zz}\neq 0, \epsdot_{yy}\neq\epsdot_{zz}
\end{array}\right.
\end{equation}

\newcommand{\eqskip}{1ex}

Finally, the solution to Eqn. (\ref{h_xz_ode}) depends on the values of all three diagonal elements of $\gradu$, yielding 15 possible solutions of which eight are volume-preserving (i.e. $\text{Tr}[\gradu]=0$):

\begin{equation}
\hspace{-6em}h_{xz}(t) = h_{xz}(0)e^{\epsdot_{xx}t} + \left\{
\begin{array}{l l}
h_{yz}(0)\gammadot_{xy}t + h_{zz}(0)\left[\gammadot_{xz}t+\frac{1}{2}\gammadot_{yz}t^2\right],
& \epsdot_{xx} = \epsdot_{yy} = \epsdot_{zz} = 0, \\
\vspace{\eqskip}\\
h_{yz}(0)\frac{\gammadot_{xy}}{\epsdot_{xx}}\left(e^{\epsdot_{xx}t}-1\right) + h_{zz}(0)\left[\left(\frac{\gammadot_{xz}}{\epsdot_{xx}}+\frac{\gammadot_{xy}\gammadot_{yz}}{\epsdot_{xx}^2}\right)\left(e^{\epsdot_{xx}t}-1\right) - \frac{\gammadot_{xy}\gammadot_{yz}}{\epsdot_{xx}}t\right],
& \epsdot_{xx}\neq 0, \epsdot_{yy} = 0, \epsdot_{zz} = 0, \\
\vspace{\eqskip}\\
h_{yz}(0)\frac{\gammadot_{xy}}{\epsdot_{yy}}\left(e^{\epsdot_{yy}t}-1\right)+h_{zz}(0)\left[\gammadot_{xz}t+\frac{\gammadot_{xy}\gammadot_{yz}}{\epsdot_{yy}}\left(\frac{e^{\epsdot_{yy}t}-1}{\epsdot_{yy}}-t\right)\right],
& \epsdot_{xx}=0, \epsdot_{yy}\neq 0, \epsdot_{zz}=0, \\
\vspace{\eqskip}\\
h_{yz}(0)\gammadot_{xy}t + h_{zz}(0)\left[\frac{\gammadot_{xz}}{\epsdot_{z}}\left(e^{\epsdot_{zz}t}-1\right) + \frac{\gammadot_{xy}\gammadot_{yz}}{\epsdot_{zz}}\left(\frac{e^{\epsdot_{zz}t}-1}{\epsdot_{zz}}-t\right)\right],
& \epsdot_{xx} = 0, \epsdot_{yy} = 0, \epsdot_{zz}\neq 0, \\
\vspace{\eqskip}\\
h_{yz}(0)\frac{\gammadot_{xy}}{\epsdot_{yy}}\left(e^{\epsdot_{yy}t}-1\right)
& \epsdot_{xx}=0, \epsdot_{yy}\neq 0, \\
\hspace{1em} + h_{zz}(0)\left[\frac{\gammadot_{xy}}{\epsdot_{zz}}\left(e^{\epsdot_{zz}t}-1\right) + \frac{\gammadot_{xy}\gammadot_{yz}}{\epsdot_{yy}}\left(te^{\epsdot_{yy}t} - \frac{e^{\epsdot_{yy}t}-1}{\epsdot_{yy}}\right)\right],
& \epsdot_{zz}\neq 0, \epsdot_{yy}=\epsdot_{zz}, \\
\vspace{\eqskip}\\
h_{yz}(0)\frac{\gammadot_{xy}}{\epsdot_{yy}}\left(e^{\epsdot_{yy}t}-1\right)
& \epsdot_{xx}=0, \epsdot_{yy}\neq 0, \\
\hspace{1em} + h_{zz}(0)\left[\frac{\gammadot_{xz}}{\epsdot_{zz}}\left(e^{\epsdot_{zz}t}-1\right) + \frac{\gammadot_{xy}\gammadot_{yz}}{\epsdot_{zz}-\epsdot_{yy}}\left(\frac{e^{\epsdot_{zz}t}-1}{\epsdot_{zz}}-\frac{e^{\epsdot_{yy}t}-1}{\epsdot_{yy}}\right)\right],
& \epsdot_{zz}\neq 0, \epsdot_{yy}\neq\epsdot_{zz},\\
\vspace{\eqskip}\\
h_{yz}(0)\frac{\gammadot_{xy}}{\epsdot_{xx}}\left(e^{\epsdot_{xx}t}-1\right) &
\epsdot_{xx}\neq 0, \epsdot_{yy}=0, \\
\hspace{1em} + h_{zz}(0)\left[\gammadot_{xz}te^{\epsdot_{xx}t} + \frac{\gammadot_{xy}\gammadot_{yz}}{\epsdot_{zz}}\left(te^{\epsdot_{xx}t}-\frac{e^{\epsdot_{xx}t}-1}{\epsdot_{xx}}\right)\right],
& \epsdot_{zz}\neq 0, \epsdot_{xx} = \epsdot_{zz},\\
\vspace{\eqskip}\\
h_{yz}(0)\frac{\gammadot_{xy}}{\epsdot_{xx}}\left(1-e^{\epsdot_{xx}t}\right) &
\epsdot_{xx}\neq 0, \epsdot_{yy}=0, \\
\hspace{1em} + h_{zz}(0)\left[\frac{\gammadot_{xz}}{\epsdot_{zz}-\epsdot_{xx}}\left(e^{\epsdot_{zz}t}-e^{\epsdot_{xx}t}\right) + \frac{\gammadot_{xy}\gammadot_{yz}}{\epsdot_{zz}}\left(\frac{e^{\epsdot_{zz}t}-e^{\epsdot_{xx}t}}{\epsdot_{zz}-\epsdot_{xx}} + \frac{1-e^{\epsdot_{xx}t}}{\epsdot_{xx}}\right)\right],
& \epsdot_{zz}\neq 0, \epsdot_{xx} \neq \epsdot_{zz},\\
\vspace{\eqskip}\\
h_{yz}(0)\gammadot_{xy}te^{\epsdot_{xx}t} &
\epsdot_{xx}\neq 0, \epsdot_{yy}\neq 0, \\
\hspace{1em} + h_{zz}(0)\left[\frac{\gammadot_{xz}}{\epsdot_{xx}}\left(e^{\epsdot_{xx}t}-1\right) + \frac{\gammadot_{xy}\gammadot_{yz}}{\epsdot_{xx}}\left(\frac{1-e^{\epsdot_{xx}t}}{\epsdot_{xx}} + te^{\epsdot_{xx}t}\right)\right],
& \epsdot_{zz}=0, \epsdot_{xx} = \epsdot_{yy},\\
\vspace{\eqskip}\\
h_{yz}(0)\frac{\gammadot_{xy}}{\epsdot_{yy}-\epsdot_{xx}}\left(e^{\epsdot_{yy}t}-e^{\epsdot_{xx}t}\right) &
\epsdot_{xx}\neq 0, \epsdot_{yy}\neq 0, \\
\hspace{1em} + h_{zz}(0)\left[\frac{\gammadot_{xz}}{\epsdot_{xx}}\left(e^{\epsdot_{xx}t}-1\right) + \frac{\gammadot_{xy}\gammadot_{yz}}{\epsdot_{yy}}\left(\frac{e^{\epsdot_{yy}t}-e^{\epsdot_{xx}t}}{\epsdot_{yy}-\epsdot_{xx}} + \frac{1-e^{\epsdot_{xx}t}}{\epsdot_{xx}}\right)\right],
& \epsdot_{zz}=0, \epsdot_{xx} \neq \epsdot_{yy},\\
\vspace{\eqskip}\\
h_{yz}(0)\gammadot_{xy}te^{\epsdot_{xx}t} &
\epsdot_{xx}\neq 0, \epsdot_{yy}=\epsdot_{xx}, \\
\hspace{1em} + h_{zz}(0)\left[\frac{\gammadot_{xz}}{\epsdot_{zz}-\epsdot_{xx}}\left(e^{\epsdot_{zz}t}-e^{\epsdot_{xx}t}\right) + \frac{\gammadot_{xy}\gammadot_{yz}}{\epsdot_{zz}-\epsdot_{yy}}\left(\frac{e^{\epsdot_{zz}t}-e^{\epsdot_{xx}t}}{\epsdot_{zz}-\epsdot_{xx}}-te^{\epsdot_{xx}t}\right)\right],
& \epsdot_{zz}\neq 0, \epsdot_{xx} \neq \epsdot_{zz},\\
\vspace{\eqskip}\\
h_{yz}(0)\frac{\gammadot_{xy}}{\epsdot_{yy}-\epsdot_{xx}}\left(e^{\epsdot_{yy}t}-e^{\epsdot_{xx}t}\right) &
\epsdot_{xx}\neq 0, \epsdot_{yy}\neq 0, \\
\hspace{1em} + h_{zz}(0)\left[\frac{\gammadot_{xz}}{\epsdot_{zz}-\epsdot_{xx}}\left(e^{\epsdot_{zz}t}-e^{\epsdot_{xx}t}\right) + \frac{\gammadot_{xy}\gammadot_{yz}}{\epsdot_{yy}-\epsdot_{xx}}\left(te^{\epsdot_{yy}t} - \frac{e^{\epsdot_{yy}t}-e^{\epsdot_{xx}t}}{\epsdot_{yy}-\epsdot_{xx}}\right)\right],
& \epsdot_{zz}=\epsdot_{yy}, \epsdot_{xx} \neq \epsdot_{yy},\\
\vspace{\eqskip}\\
h_{xy}(0)\frac{\gammadot_{xy}}{\epsdot_{yy}-\epsdot_{xx}}\left(e^{\epsdot_{yy}t}-e^{\epsdot_{xx}t}\right)
& \epsdot_{xx}\neq 0, \epsdot_{yy}\neq 0, \\
\hspace{1em} + h_{zz}(0)\left[\gammadot_{xz}te^{\epsdot_{xx}t} + \frac{\gammadot_{xy}\gammadot_{yz}}{\epsdot_{zz}-\epsdot_{yy}}\left(te^{\epsdot_{xx}t} - \frac{e^{\epsdot_{yy}t}-e^{\epsdot_{xx}t}}{\epsdot_{yy}-\epsdot_{xx}}\right)\right],
& \epsdot_{zz}=\epsdot_{xx}, \epsdot_{xx} \neq \epsdot_{yy},\\
\vspace{\eqskip}\\
h_{yz}(0)\gammadot_{xy}e^{\epsdot_{xx}t} + h_{zz}(0)\left[\gammadot_{xz}te^{\epsdot_{xx}t} + \frac{1}{2}\gammadot_{xy}\gammadot_{yz}t^2e^{\epsdot_{xx}t}\right],
& \epsdot_{xx}\neq 0, \epsdot_{yy}=\epsdot_{xx},\\
& \epsdot_{zz}=\epsdot_{xx}, \\
\vspace{\eqskip}\\
h_{yz}(0)\frac{\gammadot_{xy}}{\epsdot_{yy}-\epsdot_{xx}}\left(e^{\epsdot_{yy}t}-e^{\epsdot_{xx}t}\right)
& \epsdot_{xx}\neq 0, \epsdot_{yy}\neq 0, \\
\hspace{1em} + h_{zz}(0)\left[\frac{\gammadot_{xz}}{\epsdot_{zz}-\epsdot_{xx}}\left(e^{\epsdot_{zz}t}-e^{\epsdot_{xx}t}\right) + \right.
& \epsdot_{zz}\neq 0, \epsdot_{xx} \neq \epsdot_{yy}, \\
\hspace{1em}\left.\frac{\gammadot_{xy}\gammadot_{yz}}{\epsdot_{zz}-\epsdot_{yy}}\left(\frac{e^{\epsdot_{zz}t}-e^{\epsdot_{xx}t}}{\epsdot_{zz}-\epsdot_{xx}} - \frac{e^{\epsdot_{yy}t}-e^{\epsdot_{xx}t}}{\epsdot_{yy}-\epsdot_{xx}}\right)\right], 
& \epsdot_{xx} \neq \epsdot_{zz}, \epsdot_{yy}\neq\epsdot_{zz}.
\end{array}\right.
\label{eqn:lammps-box-update}
\end{equation}
\end{widetext}

\section*{References}
\bibliography{references}% Produces the bibliography via BibTeX.

\end{document}